\documentclass[runningheads,a4paper]{llncs}

\usepackage{amssymb}
\usepackage{cite}
\usepackage{multirow}
\usepackage{balance}
\usepackage[pdftex]{graphicx}
\graphicspath{{./figures/}{./jpeg/}} 
\DeclareGraphicsExtensions{.pdf,.jpeg,.png}

\usepackage{url}
\newcommand{\keywords}[1]{\par\addvspace\baselineskip
\noindent\keywordname\enspace\ignorespaces#1}

\begin{document}

\mainmatter  

\title{Social Politics: Agenda Setting  and Political Communication on Social Media}

\titlerunning{Agenda Setting  and Political Communication on Social Media}

%
%
\author{Xinxin Yang$^1$ \and Bo-Chiuan Chen$^1$ \and Mrinmoy Maity$^1$\and Emilio Ferrara$^2$}
\authorrunning{Agenda Setting  and Political Communication on Social Media}

\institute{$^1$Indiana University, Bloomington, IN (USA) \\ $^2$University of Southern California, Los Angeles, CA (USA) }

%
%

\maketitle

\begin{abstract}
Social media play an increasingly important role in political communication.
Various studies investigated how individuals adopt social media for political discussion, to share their views about politics and policy, or to mobilize and protest against social issues.
Yet, little attention has been devoted to the main actors of political discussions: the politicians. In this paper, we explore the topics of discussion of U.S. President Obama and the 50 U.S. State Governors using Twitter data and agenda-setting theory as a tool to describe the patterns of daily political discussion, uncovering the main topics of attention and interest of these actors. We examine over one hundred thousand tweets produced by these politicians and identify seven macro-topics of conversation, finding that Twitter represents a particularly appealing vehicle of conversation for American opposition politicians. 
We highlight the main motifs of political conversation of the two parties, discovering that Republican and Democrat Governors are more or less similarly active on Twitter but exhibit different styles of communication. 
Finally, by reconstructing the networks of occurrences of Governors' hashtags and keywords related to political issues, we observe that Republicans form a tight core, with a stronger shared agenda on many issues of discussion.
\end{abstract}

\section{Introduction}
The widespread adoption of social media is challenging the way traditional media have been used to distribute news, and to discuss top social and political issues~\cite{lerman2010information, metaxas2012social, ferrara2013traveling}.
A large body of \emph{Computational Social Science} research focuses on the study of individuals and their behaviors on such platforms~\cite{lazer2009life, boyd2012critical, pentland2014social}. Various seminal papers investigate social and political conversations on social platforms like Twitter~\cite{ratkiewicz2011detecting, effing2011social,  stieglitz2012political, bekafigo2013tweets} and Facebook~\cite{bond201261, carlisle2013social, ellison2014cultivating}. Yet, little work has been devoted to understand how the main actors of political discussion, the politician themselves, adopt and leverage such platforms~\cite{golbeck2010twitter, chi2010twitter, hemphill2013s}. 
During the 2008 Presidential Election, Barack Obama used fifteen social media sites to support his campaign. 
His successful effort demonstrated the central role of Twitter and other social platforms as integral parts of modern political communication. 
Since then, online political discussion and the attention toward political candidates and political figures, and their social media presence, arose.
Politicians are influential figures in the offline world, and surely can acquire a great deal of influence in the social media spheres as well. Their social media activity, in turn, can alter their success and affect their careers, especially during election time. 
The online campaigns preceding the 2016 Presidential Election carried out by both parties in support of various potential nominees, including Hillary Clinton, Bernie Sanders, and Donald Trump, further demonstrate the social media power to shape the political scene~\cite{wang2016deciphering, wang2016will}.
A better understanding of politicians' usage of social media channels for political conversation could therefore reveal something about the complex mechanisms of political success in the era of \emph{social politics}.

Yet, social media are not limited to political ``propaganda''.
The effects of social media political communication on the offline world are tangible. Examples of political campaigns that preceded mass mobilizations and civilian protests include the Arab Springs~\cite{howard2011opening, gerbaudo2012tweets}, Occupy Wall Street~\cite{conover2013geospatial, conover2013digital}, and the Gezi Park protest~\cite{varol2014evolution}. Although it is difficult to establish a causality link, we can safely say that the ``Twittersphere'' can be a strong indicator of political and public opinion~\cite{tumasjan2010election}.
The open nature of Twitter\footnote{At least with respect to other platforms like Facebook where ties are mostly formed based on pre-existing offline connections~\cite{demeo2014on}.} probably contributed to determine its \emph{political communicative power}. The ability to communicate interesting political issues yields the opportunity to users to acquire more visibility and influence~\cite{cha2010measuring, bakshy2011everyone, parmelee2011politics}, although Twitter political discussion is plagued by a number of issues related to manipulation and abuse~\cite{ratkiewicz2011detecting, ferrara2014rise, ferrara2015manipulation}.

In this paper we explore how the main actors of political discussion, the politicians, adopt Twitter to cover social and political issues. We focus on U.S. President Obama and all the 50 U.S. State Governors, and adopt the framework of \emph{agenda-setting theory} to identify their main topics of discussion.
The analysis of over one hundred thousand of their tweets  reveals how Governors and the President use Twitter, what are the emerging patterns of political discussion, the top issues for each party, and finally who are the politicians who exhibit the most coherent political agenda.

\section{Social Media and Politics}
Twitter was born in 2006. In less than 10 years, it acquired half billion users, 310 million of which are active and produce over 500 million tweets per day as of July 2016.\footnote{Twitter official data: \url{https://about.twitter.com/company}} Twitter suggests that ``each tweet represents an opportunity to show one's voice and strengthen relationships with one's followers".\footnote{Twitter official blog: \url{https://blog.twitter.com/2014/what-fuels-a-tweets-engagement}}
As a modern political toolbox, Twitter has been widely used by various Presidents, Congressmen, Governors, and other politicians all over the world. In particular in the United States, Twitter and other social media have been not only the subject of extensive research, but also the platforms used to run large-scale social experiments to study political mobilization~\cite{bond201261}.
Scholars from various disciplines have investigated the role of these platforms in modern political communication. 

Generally, social media research related to politics can be categorized into two fields. The former focuses on the possibility of using social media signals to predict political elections. A large number of papers faced this challenging question, with at times promising results.  For example, Gibson and McAllister's  study~\cite{gibson2006does} demonstrated a significant relationship between online campaigning and candidate support. Macnamara found evidence of a ``significant online political engagement'' in the 2008 U.S. Presidential Election~\cite{macnamara2010quadrivium}. Other studies covered the U.S. Presidential debate and Twitter sentiment, finding an alignment between popular opinions and votes~\cite{diakopoulos2010characterizing, stieglitz2012political, digrazia2013more}. Despite some promising work, the issue of predicting elections using social data remains debated~\cite{gayo2012wanted}.

The second area of research investigates Twitter users' behaviors, opinions and topics of political interest, at times proposing methods to identify their political alignments~\cite{conover2011predicting, cohen2013classifying}. Some of these studies highlighted interesting socio-political phenomena: for example, Conover \emph{et al.}~\cite{conover2011political} found that the network of political retweets exhibits a highly segregated bipartisan structure, which seems to reflect the users' political leanings, similarly to political blogs~\cite{adamic2005political}.
Shogan's \textit{et al.} research showed that, in recent years, Republican politicians tweeted more than five times as often as Democrats, suggesting that Twitter might be particularly appealing to American opposition politicians, who use it as an instrument for voicing their dissent directly to the public~\cite{glassman2009social, shogan2010blackberries}. A study conducted by Chi and Yang~\cite{chi2010twitter} found that Democratic congressmen tend to release information that citizens want to hear, while Republican congressmen share with the citizens their own agenda. Hemphill's work suggested that Congressmen of opposing parties use very different strategies to choose the hashtags that better reflect their framing efforts~\cite{hemphill2013framing}.

It appears that most literature either focuses on Twitter and elections, especially before and during election time, or focuses on President or Congressmen, even though ``most Americans have more daily contacts with their state and local governments than with the federal government''.\footnote{White House: State and Local Government, 2015 \url{https://www.whitehouse.gov/1600/state-and-local-government}}

Studies on State Governors and their social media presence are absent, and this paper aims at filling this gap. Although some research focuses on how politicians use social media before and during their election, what happens after that? Voters are excited about their party's success, and they are vocal about it. What comes after this initial excitement? We want to shed light on which Governors really follow their agenda after their election, and determine whether a framing of clear intents and goals emerges from their political channels online.

As of April 25, 2015, the 50 U.S. State Governors in charge collectively gathered over 3  million followers and sent out over 150,000 tweets. Though the majority of their Twitter accounts are merely political, some, such as Michigan Governor Rick Snyder's ``OneToughNerd'' account, show some character's personality traits, while others lend a certain intimacy, for example including family pictures like for Maryland Governor Larry Hogan, New Jersey Governor Chris Christie, Maine Governor Paul LePage and Louisiana Bobby Jingdal. Balancing personal lives and public service information makes State Governors' Twitter accounts very interesting objects to study the Governors' political stance in front of the public. This paper tries to dig into this unexplored field to analyze the State Governors' Twitter accounts by using agenda-setting theory, to understand whether the State Governors' activity on Twitter can be used to predict the popularity of parties or coalitions.

\section{Agenda-Setting Theory}
Twitter allows politicians to set their political agenda and reach their audience directly. Studying their behaviors brings the promising opportunity to further our understanding of \emph{agenda setting} in digital media~\cite{russell2014dynamics}. The agenda-setting theory is regarded as a key element to explain mass communication effects and mass media influence in long-term conditions. The primary assumptions of the theory were formulated by Maxwell McCombs and Donald Shaw in 1972~\cite{mccombs1972agenda}. Agenda setting is one of the most widely used theories in communication studies since then~\cite{iyengar1994anyone, iyengar2000new, weaver2004agenda, mccombs2005look, wanta2007effects}.

Agenda setting is the filter mass media perform when selecting certain issues and portraying them frequently and prominently, which leads people to perceive those issues as more important than others. Two levels of agenda-setting theory will be used in this study. The first-level agenda setting focuses on the amount of coverage of an issue, suggesting which issues the public will be more likely to be exposed to. The second-level agenda setting, also called \emph{framing} as suggested by McCombs, Shaw and Weaver~\cite{mccombs1997communication}, examines the influence of attribute salience, or the properties, qualities, characteristics, and relations. By making some political issues salient, agenda setting makes these specific issues more accessible than others.

The first level of agenda setting is the issue level. Though some scholars categorize top issues manually~\cite{russell2014dynamics}, we plan to use top issues listed on the White House's homepage. As of April 2015, the top seven issues listed were: economy, education, foreign policy, health care, immigration, climate change, energy and environment, and civil rights. April 2015 is also the time of our Twitter data collection. We will try to identify whether politicians give attention to these issues by analyzing how often kewords and hashtags related to these issue are mentioned on their Twitter accounts. 
In the second level of agenda setting, we will analyze whether Democrats and Republicans highlight different attributes of the same issue by examining the hashtags and keywords they choose when they do discuss an issue. 
We will also examine those hashtags and keywords relations by constructing occurrence networks to see how those hashtags and keywords are framed in the Governors' tweets.

Many researchers found different tweeting patterns among Democrats and Republicans Congressman, such as Shogan \textit{et al}.~\cite{glassman2009social, shogan2010blackberries} and Chi and Yang~\cite{chi2010twitter}. Our research as well aims to find whether State Governors' Twitter accounts exhibit different levels of engagement. 
Then, we would like to further our understanding of the general patterns of usage, applying the second level agenda-setting theory, or framing, to scrutinize the hashtags and keywords network structure. Hence, we formalize the following three research questions:

\textbf{RQ 1}: How frequently do Governors use Twitter to discuss their political agenda? Do party differences emerge?

\textbf{RQ 2}: How do Governors' Twitter accounts reflect their political agenda, and how similar political agendas are across Governors?

\textbf{RQ 3}: What similarities and differences emerge in hashtag usage among Governors' Twitter accounts?

\section{Data Collection}
We used the Governors' timelines to reference the tweets from the 50 U.S. Governors and the U.S. President Barack Obama. We collected 114,316 tweets from the Governors' timelines. We downloaded the stream of tweets for each account by querying the Twitter Public API for user timeline by using a manually-collected list of account names. This returns the entire stream of tweets for each account, avoiding sampling issues~\cite{morstatter2013sample}. We performed the queries between January 23 and April 26, 2015, for all 51 accounts, in a systematic way and with a 100 second pause between each account. The pause was set to prevent our script from sending queries that exceed the rate limitation of the API. All data were finally stored into a JSON file and later analyzed.

We parsed each tweet to extract words and hashtags using the regular expression package \textit{re} with Python 3. We first removed the URLs by excluding patterns starting with http, https, ftp, and mailto. Then, tweet texts were converted into lowercase for consistency. Finally, we obtained hashtags and words by another set of regular expressions. The hashtags were defined as sets of concatenated characters starting with a pound sign (\#), while the words were defined as concatenated sets that start and end with alpha-numeric characters.

We identified the keywords by manually looking for the most frequent words that could be indicative of specific topics and sound meaningful to ordinary readers. To identify what could be the candidate words associated with each topic, we first manually parsed our collection of tweets and assigned the words that appeared together with the target topic as the candidate word selection for that topic.\footnote{Given the massive size of the dataset, with over one hundred thousand tweets, this procedure required three annotators and countless hours of work.}
For example, when we query for ``health care'' we will assign each of the 17 words (we, will, fight, to, protect, the, healthcare, of, Floridians, their, right, to, be, free, from, federal, overreach) appearing in the tweet ``We will fight to protect the healthcare of Floridians \& their right to be free from federal overreach.'' as a candidate choice of keywords for health care. All the stop-words that were identified by the Python Natural Language Toolkit (NLTK) were removed. In the previous example, the set of candidate words after this further cleansing is reduced to (fight, protect, heathcare, Floridians, right, free, federal, overreach). The next step was to remove the words that are syntactically needed but not contextually meaningful. We identified the words that were a keywords of more than one topic and manually marked them to be further removed or not. Words that were shared by more than one topic were marked to be deleted if we were unable to find a potential topic for them; words that possibly related to any of the topics were marked to be kept. In the example, words to be deleted included: fight, protect, Floridians, right, federal, overreach. These words could not be attached unequivocally to any one topic. For example, the words \textit{fight} and \textit{protect} appeared more often attached to foreign and immigration issues, and the word \emph{right} appeared more often related to civil right issues. Words to be kept included: \textit{healthcare} (as well as \textit{health care} with a space), and \textit{freedom}, which could be assigned to health care, in particular related to the Affordable Care Act (or, ObamaCare). After we identified which words to delete or keep, we then updated the sets of each candidate keywords for each topic. We then ranked each candidate keywords by their overall frequency in our collection. The top seven candidate keywords for each category were used to identify the topic of each tweet. We assigned a tweet to a topic whenever any of the 7 keywords for a topic appeared in a tweet. The topics were not mutually exclusive: in other words, one tweet could be assigned to more than one topic when the top candidate keywords from different categories occurred in a tweet. We counted the numbers of tweets for each topic among the Governors. The agenda was finally recovered by ranking the topics by the numbers of tweets associated to them: the results are displayed in Table~\ref{tab:top_words}. The assessment of the quality of the agenda produced by our semi-automatic method is satisfactory: the seven topics are each clearly identified by a short list of intuitive keywords. By means of the same approach, we varied the number of keywords to include more words, finding that the results (discussed later) were substantially unaltered. Finally, the proposed method to generate the agenda was preferred over traditional topic modeling techniques that we tested, such as LDA, because of the inability of such probabilistic generative models~\cite{blei2012probabilistic} to discriminate between topics related to issues relevant to politics, and other irrelevant (for our purpose) topics that appeared in the Governors' Twitter timelines.

\begin{table*}[!t]
\centering 
\caption{Top Words per Category}
 \begin{tabular}{@{}c@{} c@{} c@{} c@{} c@{} c@{} c@{}}
 Civil Right & Economy & Education & Energy and  & Foreign & Health Care & Immigration \\ [0.8ex] 
 & & & Environment & & & \\
 \hline\hline
 veterans & economic & education & energy & drug & health & investments\\ 
 \hline
 citizens & economy & students & manufacturing & sexual & food & immigration\\
 \hline
 rights & unemployment & school & water & assault & medicaid & employment\\
 \hline
 equal & manufacturing & veterans & affordable & campuses & insurance & sustainable\\
 \hline
 marriage & employees & schools & climate & uniform & transportation & struggling\\ 
 \hline
  defense & transportation & kids & tech & foreign & affordable & action\\
 \hline
  restoration & companies & college & capital & asia & freedom & portfolio\\
 \hline
\end{tabular}
\label{tab:top_words} 
\end{table*}

\section{Experimental Results}

\subsection{Overall Tweeting Patterns}
To try answer \textbf{RQ 1}, we analyze the 114,316 tweets collected from the Governor's timelines. The amount of tweets produced by each Governor ranged from 30 to 3,242, with a median of 2,838.
These figures demonstrate that the majority of Governors is quite active on the platform. There were 46,125 tweets posted by the 19 Democrat Governors, and 68,047 by the 30 Republican ones: this suggests that, on average, each Democrat produced 2,427 tweets, and each Republican posted 2268 tweets; this difference is not particularly significant.
President Obama contributed 3,242 to the Democrats, and the independent Governor of Arkansas had 144 tweets.  
We were able to identify 75,202 hashtags and words from the tweet texts after removing the URLs. Democrat Governors used 50,960 words while Republican governors used 41,263. 
The Democrats also tweeted more distinct hashtags, 6,463, while Republicans had only 4,264. 
A previous study conducted by Shogan \textit{et al}.~\cite{glassman2009social, shogan2010blackberries} on the House  tweeting patterns suggested that Republicans tweet more, and Twitter might be particularly appealing to the American opposition politicians. 
Our analysis demonstrates that there are no significant differences in terms of average posting volumes between the two parties, and the larger sheer number of Republican tweets is to be attributed to the significantly greater number of Republican Governors (30 versus 19 Democrats).
However some stylistic differences emerge, in that Democrat Governors seem to make a much more pervasive and diverse use of hashtags than Republicans.

\subsection{Political agenda and keywords usage}

To answer \textbf{RQ 2}, we plan to describe each Governor's posting behavior according to the agenda we defined in Table~\ref{tab:top_words}. For each Governor's account, we calculated the number of times each keyword of Table~\ref{tab:top_words} appeared in any of the Governor's tweets. By sorting this dictionary of keywords and relative usage in descending order, we can obtain a rank of each Governor's keyword usage. We can therefore use the ranked keyword dictionaries to perform pairwise comparisons of Governors and try capture similarities and differences in priorities regarding the categories of political discussion. Note that using rankings is preferable to using simple feature vectors of  keyword counts: ranks are more amenable to direct comparisons (for example via Spearman's rank correlation) without data normalization to account for different intensity of activities and other biases.

\begin{figure}[t!]  \centering
  \caption{Distribution of Spearman Rank Correlation scores}
  \includegraphics[width=\columnwidth]{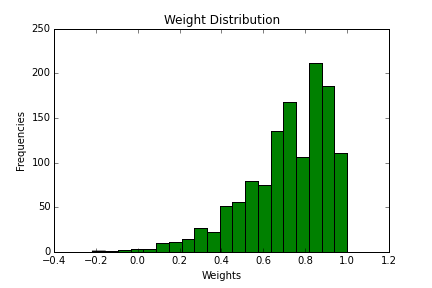}
  \label{fig:spearman}\vspace*{-.75cm}
\end{figure}

To measure the correlation of discussion keywords between all pairs of Governors, we use Spearman's correlation applied on their ranked keyword dictionaries. Spearman's rank correlation assigns each pair $<X_i, X_j>$  a similarity score between -1 and 1, with $X_i$ and $X_j$ being the keyword ranks of Governors $i$ and $j$ respectively. Score of 1 and -1 indicate perfect positive and negative correlation, respectively, whereas a score of 0 suggests no correlation.
To understand the distribution of pairwise correlation scores, we plotted Figure~\ref{fig:spearman}. The range of scores spans roughly from $-0.2$ (indicating a slight negative correlation) to very strong positive correlation scores greater than 0.8. The skewness towards positive scores can be attributed to the fact that we have considered only seven words per category, with seven total categories, for determining the rank distributions.  

\begin{figure}[t!]  \centering
  \caption{Keyword-based correlation among Governors}
  \includegraphics[width=\columnwidth]{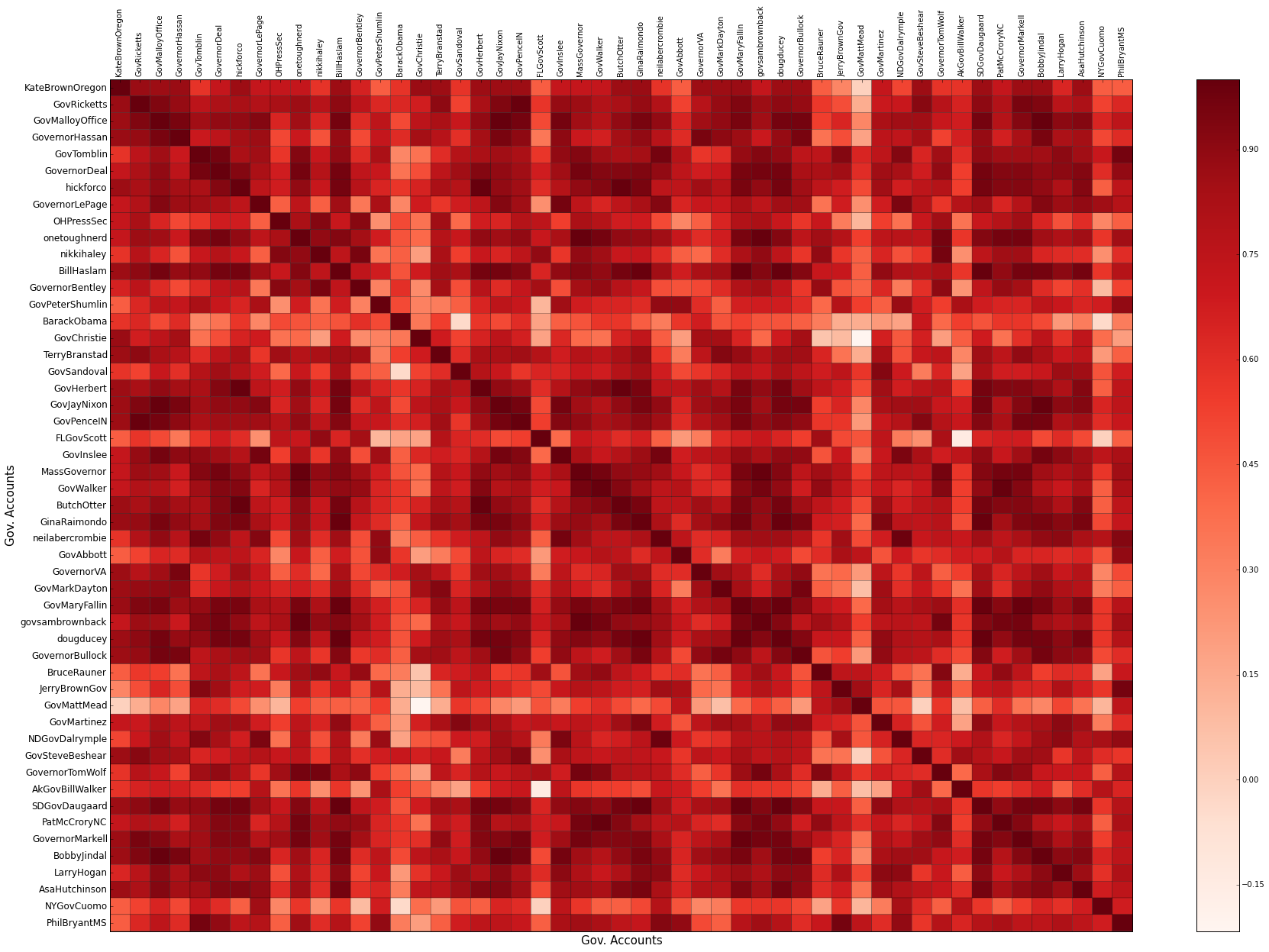}
  \label{fig:matrix}\vspace*{-.75cm}
\end{figure}

\begin{figure}[t!] \centering
  \caption{Governors Network through the lens of Agenda Setting Theory}
  \includegraphics[width=\columnwidth,clip=true,trim=160 80 220 80]{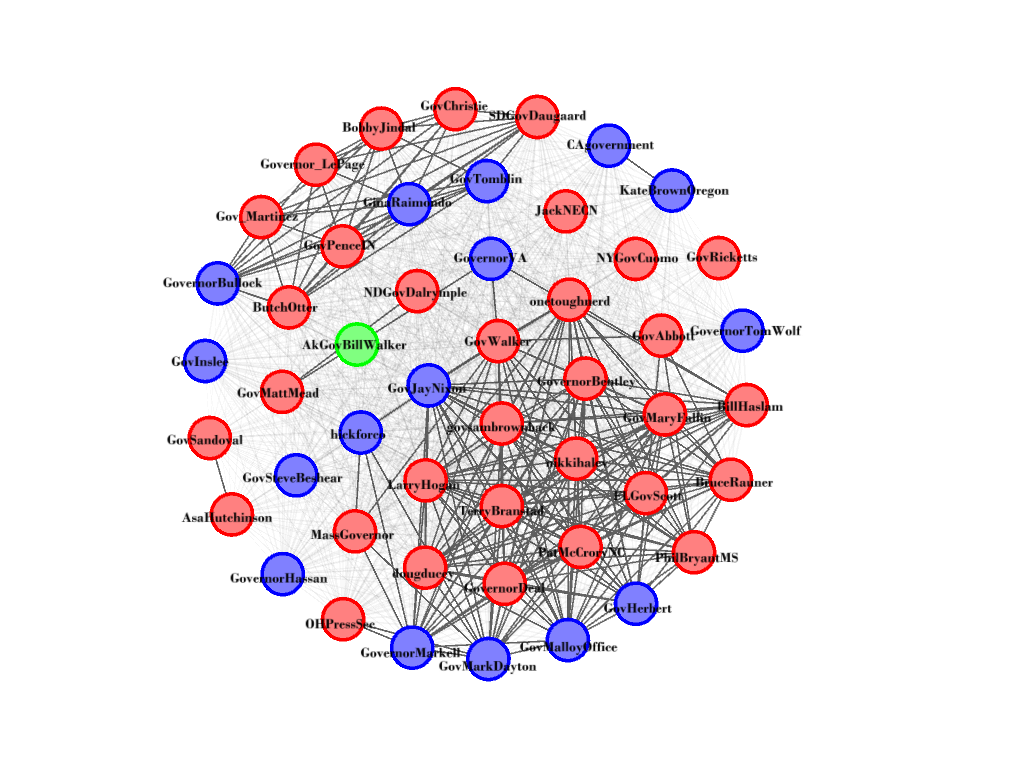}
  \label{fig:governor_network}\vspace*{-.75cm}
\end{figure}

Figure~\ref{fig:matrix} shows the matrix of pairwise Spearman correlations among the 50 U.S. Governors plus the U.S. President Barack Obama. The visual inspection of Figure~\ref{fig:matrix} suggests the presence of a strong block structure, as groups of highly correlated accounts happen to be clearly identifiable. To further inspect this hypothesis, we generated a weighted graph of inter-Governor similarity using the matrix of Figure~\ref{fig:matrix} as adjacency matrix. The resulting graph is displayed in Figure~\ref{fig:governor_network}, where for visual clarity,  self-loops have been removed and all edges with weights (i.e., Spearman correlation) less than 0.8 have been filtered out. 
Figure~\ref{fig:governor_network} captures the agenda similarity network among Governors.
Its analysis suggests the emergence of a strong community structure, where Republican Governors appear to be strongly aligned on agenda priorities and form two tight clusters: the larger red cluster revolves around Wisconsin Governor Scott Walker, California's Jerry Brown, Maryland's Larry Hogan, Iowa's Terry Branstad and few others. 
The second red cluster revolves around (former) Indiana Governor (and current Vice President nominee) Mike Pence, Maine's Paul LePage, New Jersey's Chris Christie, Luisiana's Bobby Jindal and few others.

The similarity, in terms of agenda priorities (as measured by the rank correlation) seems to be much less pronounced for Democrats: Governor of Rhode Island Gina Raimondo and West Virginia's Earl Tomblin, and Virginia's Terry McAuliffe form a small central cluster, whereas Colorado Governor John Hickenlooper, Missouri's Jay Nixon, Kentucky's Steve Beshear, and New Hampshire's Maggie Hassan show some agenda similarity. All the other Democrats Governors somehow sit at the periphery of this network showing spurious alignments with some of their Republican counterparts, and a less pronounces inter-party agenda priority sharing.

\subsection{The Governor-hashtag graph}
\begin{figure}[t!] \centering
\caption{Governors and Hashtag network} 
\includegraphics[width=\columnwidth,clip=true,trim=0 140 0 140]{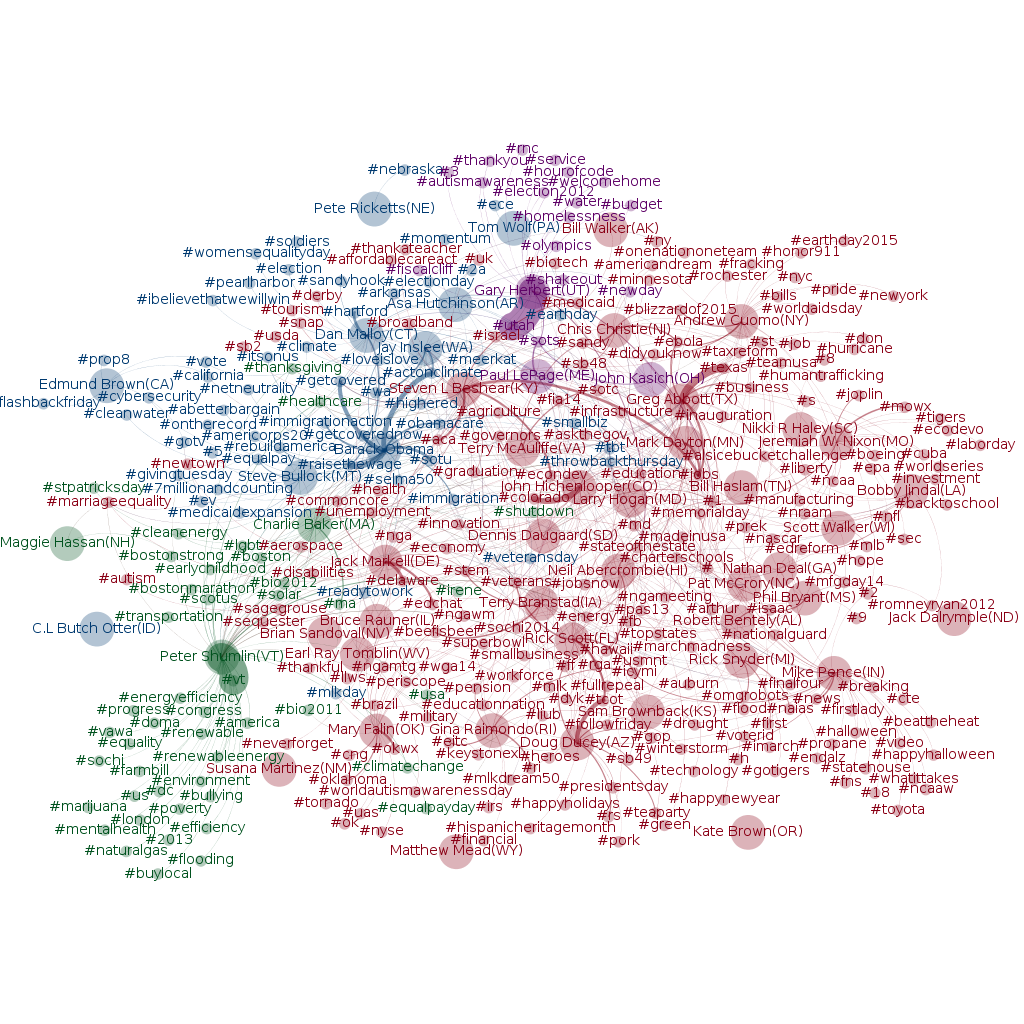}
\label{fig:governors_hashtag_network}\vspace*{-.75cm}
\end{figure}

To address \textbf{RQ 3}, we finally explored the similarity among the governors at a hashtag level. We extracted the hashtags from each Governor's timeline and created a Governor-hashtag graph. The nodes in this bipartite graph represent the Governors and the hashtags they used. A Governor node and a hashtag node would be connected if the Governor had used the hashtag in any of his/her tweet. The weight is the number of tweets that contain that hashtag. We only extracted the hashtags that were used more than 10 times among all the Governors and by more than two Governors, to focus specifically on more common hashtags.
We were able to identify 658 common hashtags that occurred more than 10 times and were used by more than two Governors from our collection.  
We also tried to recover the community structure by  using the Louvain modularity maximization algorithm~\cite{blondel2008fast}.
The result for the Governors' hashtag usage are demonstrated in Figure~\ref{fig:governors_hashtag_network}. 
The graph only represents the nodes that were connected with edges with weights larger than four, for visual clarity. The large circles denoted the nodes for Governors, and the small ones were nodes for hashtags.
 
We were able to identify four communities using the modularity algorithm with the resolution set to $2.0$. Varying the resolution limit parameter~\cite{fortunato2007resolution} provided consistent results. The four communities contained 36, 9, 3 and 3 Governors, respectively, as shown in Figure~\ref{fig:governors_hashtag_network}. We colored the largest community in red to indicate that it's the community with the largest fraction of Republican Governors (24). The second community is colored in blue to indicate that it's the community with the largest fraction of Democrats (8). The other two communities were colored in green and purple, respectively. We believe that the green cluster should belong to the Democrats (it contains Dems like Vermont's Peter Shumlin and New Hampshire's Maggie Hassan); the purple cluster contains several Republican Governors (e.g., Ohio's John Kasich and Maine's Paul LePage). Overall, the clustering algorithm assignment was correct for 32 of the 51 Governors (62.7\%). It generated 24 correct assignments out of the 30 Republicans (80.0\%), 8 correct among the 19 Democrats (42.1\%), and the Independent Governor of Arkansas was assigned to the reds. 

In light of the most meaningful keywords for each of the seven categories summarized by Table~\ref{tab:top_words}, we parsed each Governor's timeline to determine to what extent the tweets of each individual were representative of each category. The underlying assumption of this strategy is that the more a State Governor tweets about any particular category, the more he/she is concerned about that particular issue, or at least wants to convey that message to his/her followers. 
In general, for both parties, it is quite easy to scrutinize the most recurring topics of discussion of each Governor and identify those who concentrate more or less on politics and policy related topics, or or other types of events.

Figure~\ref{fig:governors_hashtag_network} illustrates the most commonly occurring hashtags and issues of discussion of the two groups. Its analysis yields a good amount of insights into U.S. political discussion. One can notice the commitment of certain Governors to specific topics: for example, Vermont Governor Peter Shumlin seems pushing an agenda focused on environment, energy, and local economy issues. Other Democrats, like Connecticut Governor Dan Malloy, Arkansas' Asa Huthinson, the U.S. President Obama, focus on issues related to climate change, equality, health care, and education.

The Republican agenda is sufficiently diverse but focuses mostly on issues related to economy (small business, innovation, ``made in USA'', agriculture), immigration and security (human trafficking, Texas), and civil rights (especially veterans', military, and marriage rights).
A number of external events are also discussed (note that we did not remove any hashtags from the Governor-hashtag graph as long as it matched the threshold criteria explained above): some examples include reference to sport events (Nascar, Basket's March Madness, etc.), political events (2012 Elections, the GOP Convention, etc.) and tragedies (the Boston Marathon bombing, the Sandy Hook school shooting, etc.).

\section{Conclusions}
In this article we explored the landscape of U.S. Governors political communication on Twitter using the tool of agenda setting theory. We first collected a sizable amount of tweets (over one hundred thousand) generated by these politicians, and assessed that most of them are quite active Twitter users. Our results clarified some previous research about the usage of social media platforms by Democratic and Republican politicians, showing that Republican and Democrat Governors tend to be more or less equally active on Twitter on average, however they exhibit different styles of communication, with the Democrats significantly more inclined to use hashtags than their counterparts. 

We furthered our understanding of Governors' priorities using the agenda-setting theory to identify a set of seven categories of top socio-political issues, by means of a semi-automatic annotation strategy.
After inferring the priorities of each Governor, and computing the pairwise similarity among Governors, we constructed a network that reflects Governor agendas similarity. Its analysis illustrates that President Obama has a distinctive agenda-setting strategy, which has no affinity with either Democrats or Republicans. 

The graph also shows that Republican Governors, such as Arkansas Governor Asa Hutchinson, Tennessee's Bill Haslam,  Louisiana's Bobby Jindal, Oklahoma's Mary Fallin, and Arizona's Gough Ducey, shared the most similar issue agenda settings. Republican Governors, compared to Democratic Governors, tend to be more clustered, which suggested that they share their party issues and priorities more than their Democrat counterparts, and in turn these priorities can be easily identified through the Governors' messages. 
Many Democratic Governors did not align significantly with their party colleagues in top issue agendas. This suggested that Democrats have less cohesive priorities on Twitter as compared their Republicans counterparts. Similar insights were brought by examining the  Governor-hashtag graph we built from hashtag usage patterns. 

This study displayed the high-level dynamics of adoption of Twitter by U.S. Governors based on how they set their agenda on top political issues and how they frame their conversation around it. Further studies should explore the \textit{public agenda setting}, which means the agenda setting of the public in each State, to see if these share similar trends with their Governors' agendas. This would shed light on the effects of politicians' social media conversation on the public.

\bibliographystyle{splncs03}
\bibliography{COSN}

\end{document}